\documentstyle{article}

\begin{document}
\title{Breaking and Splitting asteroids by nuclear explosions
        to propel and deflect their trajectories}
\author{
D. Fargion, \\ Physics Dept. Rome University 1, Rome; \\
            I.N.F.N., Rome; \\
            Technion Institute, Physics Dept. Haifa, Israel}
\date{Rome 19/03/1998}
\maketitle
\begin{abstract}
Splitting by atomic bombs an asteroid in flight is the best way to
deflect its trajectory. How and when it should be done is described.
\end{abstract}
We might need , in a near or late future to deviate incoming asteroid
\cite{ref_1} as the approaching 1997XF11, at maximal deflection angle
for any given input energy E, energy which may be due to energetic nuclear explosions.
If the explosion  shock wave will hit near the asteroid surface
their blast will deposit by pure
radiation pressure only a small momentum: $ \Delta P_\perp = E/c$,
leading to a negligible deviation angle on the asteroid trajectory:
\begin{equation} \label{eq1}
\Delta \theta = \frac{\Delta P_\perp}{P} = \frac{E}{M \,V_o \,c} =
10^{-11} \left( \frac{E}{20\;MT} \right) \left( \frac{V_0}{30 \; km \; s^{-1}}
\right)^{-1}
\left(\frac{M}{10^{16}\, g } \right)^{-1}
\end{equation}
where M and $V_0$ are the asteroid mass and present velocity at characteristic
values and the thermonuclear energy unity, i.e. tens of Megatons.
This poor deviation angle, even if applied now, will be by many orders
of magnitude too small to move the final trajectory far away from the Earth
target. On the contrary the same amount of energy (nearly a thousand Hiroshima
bomb energy), while splitting the body, may offer a "sufficient kinetic"
kick to deflect the main asteroid
body and its secondary massive fragment. The bomb must explode on the asteroid
or, better, inside it and it must expel a fragment of mass $m_1$. The large
efficiency in deviate the asteroid breaking its body derive by simple
Newtonian cinematic formula below. \\
Let us label respectively the main asteroid
mass and velocity (toward  Earth) (M, $V_0$) and its fragment mass and velocity
(in the center of mass, c.m., system) ($m_1$, $v_1$) and, finally, the main relic
asteroid body and velocity (also in c.m. system) ( $\left( M - m_1 \right)$, $v_2$ ).
Assuming an
efficient conversion of the nuclear blast in kinetic energy in breaking the
asteroid of the order of unity, one finds at general relativistic and non-relativistic
regimes:
\begin{eqnarray}\label{eq2}
  E & = & m_1 \, c^2 \left( \gamma_1  - 1 \right) + \left( M - m_1 \right)
  c^2 \left( \gamma_2 - 1 \right) \simeq \nonumber \\
  & & \simeq \frac{1}{2} m_1 \, {v_1}^2
  \, + \, \frac{1}{2} \left( M - m_1 \right) \, {v_2}^2
\end{eqnarray}
%
%
\begin{equation}\label{eq2_b}
  E \simeq \frac{1}{2}
  \frac{M}{m_1} \left( M - m_1 \right) \, {v_2}^2
\end{equation}
when $m_1$ is smaller than $E / c^2$ then the relativistic
equation applies and one gets the same result as in equation
\ref{eq1}; otherwise ( $ \left( M - m_1 \right) \gg m_1 \gg \, E\,/\,c^2 $ )
the non-relativistic approximation holds and we apply
momentum conservation for equation \ref{eq2_b}. Consequently the deviation angle for the
main relic object (if the applied momentum is orthogonal to the primordial one)
is:

\begin{eqnarray}\label{eq3}
  \tan \Delta \theta_2 & = & \; \frac{v_2}{V_0} \; = \;
  \frac{\sqrt{2\,m_1\,E}}{\sqrt{M\,\left( M - m_1 \right)}\,V_0}
  \simeq \nonumber \\
  & \simeq &  4.5 \cdot 10^{-5} \left( \frac{m_1}{10^{12}g} \right)^{\frac{1}{2}}
  \left( \frac{M}{10^{16}\,g} \right)^{-1}
  \left( \frac{V_0}{30 \, km \, s^{-1}} \right)^{-1}
  \left( \frac{E}{20\,MT} \right)^{\frac{1}{2}} \\
  & & \nonumber \\ \label{eq3_b}
  \tan \Delta \theta_1 & = & \frac{v_1}{V_0} =
  \frac{\sqrt{2\,\left( M - m_1 \right)\,E}}{\sqrt{M\,m_1}\,V_0}
  \simeq \nonumber \\
  & \simeq & 0.3 \left( \frac{m_1}{10^{12}\,g} \right)^{- \frac{1}{2}}
  \left( \frac{E}{20\,MT} \right)^{\frac{1}{2}}
  \left( \frac{V_0}{30 \, km \, s^{-1}} \right)^{-1}
\end{eqnarray}
%
where $\Delta \theta_1$ and $\Delta \theta_2$ are the deflection
angles of the vectors $v_1$ and $v_2$ with respect to the incoming direction
of the asteroid;
we assumed a fragmentation mass $m_1$ corresponding to a hundred meter volume,
comparable to a small nuclear bomb \mbox{"crater".}
The above deflecting angle $\Delta \theta_2$ , while small, is
large enough to divert the trajectory of the bolid at a reasonable  impact
parameter, either if acted now or even at a quarter of its present distance.
Indeed for a quarter of distance of the "dangerous" 1997XF11,
$D/4 \simeq - \frac{26}{4} y \cdot V_0 \simeq 2 \cdot 10^4 sec \cdot c$,
a Jupiter like
distance, one derives the final impact parameter distance $\Delta
b$:
\begin{equation}\label{eq4}
  \Delta b \; \simeq D \; \Delta \; \theta_2 \; \approx \; 40 \; R_\oplus
\end{equation}
i.e. a distance which is comparable with a (save) Moon distance.
Let us note that $\Delta \theta_2$ is nearly five million times
larger than $\Delta \theta$ in equation \ref{eq1}: therefore no
doubt that the fragmentation process is a key propelling mechanism in
deflecting asteroid trajectories.
Technical problems to dig at deep (tens, hundred meter) a nuclear bomb
arises.\\

A "cooperative" group of explosions at a "corona" of few hundred meters, may
"bite" and "cut" an asteroid piece at better way; however synchronization of
instruments of such distances may be problematic (even if multiple nuclear
heads have been widely developed).
A "land-off" of many nuclear heads and their coherent explosion may be the best
solution. From equation \ref{eq3} one notices that the deflection angle by
fragmentation grows only by the square root of the nuclear energy. Therefore
four times (E/4) energetic bombs in "coherent" explosions, may do a total
deflection twice larger than an unique one of the same total energy E. Therefore
in principle a sequence of mini explosions may kick gently and more efficiently
the asteroid path. However an open problem is to blast in the correct way (or
side) the bombs in phase with their previous vectorial momentum kicks. This problems
may be within technological solution. Finally an asymmetric asteroid may offer
protuberances to be easily broken by nuclear bombs.\\

Let us notice that as the time grows also the
needed deflection  angle increases as shown in equation \ref{eq3}, calling for a
fragmentation energy which increases quadratically with time. For this reason
the urgency of the present article related to the \mbox{1997 XF11}
incoming which is submitted to the scientific
attention. The maximal deflection reachable occurs (from equation \ref{eq3})
when $\Delta \theta_1 \simeq \Delta \theta_2 $, i. e., when there
is enough energy to split in two equal parts the asteroid. Then
the splitting angle is:
\begin{eqnarray}
  \tan \Delta \theta_{max} & \simeq & \sqrt{\frac{2\,E\,}{M\,{V_0}^2}}
  = \nonumber \\
  & = & 4.5 \cdot 10^{-3}
    \left( \frac{M}{10^{16}\,g} \right)^{-\frac{1}{2}}
  \left( \frac{V_0}{30 \; km \; s^{-1}} \right)^{-1}
  \left( \frac{E}{20\;MT} \right)^{\frac{1}{2}} \label{eq_u}
\end{eqnarray}
The above value is hundred million times better than the
deflection angle in equation \ref{eq1}. Energetic bounds on the
energy needed to break at half a $Km^3$-size asteroid are small
(on the contrary the energy needed to "evaporate" a $Km^3$-size mountain
are severe, i.e. $E \gg 20\,MT$) but the cut and split in half may
be technically difficult and dangerous. It is auspicable that a
half asteroid would be sent towards the Sun, which may keep "clean"
in future our Solar System.
There are moreover interesting consequences associated with the
nuclear fragmentation of asteroids:
1) Let us remind the fragments problems
of Patriots rockets during the Gulf-War.
A secondary fragment while thousands of times smaller than the main
asteroid is widely deflected by angle $\Delta \theta_1$ in equation \ref{eq3_b},
with no secondary danger.
2) If one speculate on the possibilities that intelligent life ever existed in
the past in our solar system, and if it had to face similar problems finding
analogous conclusions (nuclear bombs to kick asteroids), than we may expect:
(a) some asteroids may be still radioactive or chemically polluted by past
explosions.
(b) their puzzling morphology may reflect their past explosive kicks. In this
speculative framework we remind the recent discovery of "gruviera" like
asteroids objects.\\

Moreover we speculate that (c) the asteroids may tell us on their
surface nature. Indeed very recent observations \cite{ref_1}
inform us on two possible distint population (neutral colour and
redder colour asteroids) of Kuiper-belt asteroid objects (KBOs).
These two group have not correlation between colour and any
physical or orbital parameter. Such asteroidal bimodality may
testimony that some asteroids have been heated and melted. May
really the dark ones be relics of such nuclear explosions and
deflections?\\

In conclusion we notice that a "slow" incoming asteroid
($v_F \leq 10 Km/sec$) may be also captured in an
orbit around the Earth. If it will be co-rotating with our planet it may find
a stable orbit (like our Moon), while in case of contro-rotating trajectory
it may be slowly attracted (by tidal or "Riemann" forces) towards Earth;
longer times may offer a proper solution to face with "Damocles knife" hanging
over our children's  heads.
The steady presence of an very heavy orbiting massive asteroids
(mini Moon of hundred $Kms$ size) in a bound trajectory
around Earth may influence and disturb life by periodic mini tidal
waves.
Moreover nearby encounters by visiting planets (or miniplanets) may induce severe tidal strains , by
responsable of past mass extinctions
observed on the Earth past \cite{ref_2}.
The captured asteroid destruction by tidal forces may lead to the appearance of
a final spinning ring around our planet at smaller scales as the Jovan
or Saturnine rings. In conclusion  in  order to avoid a unique death-life
 experiment it may be worthfull to test the ability in deflecting
 asteroids by splitting their mass on  other objects (small scales
 asteroids) at save trajectories and distances. Finally the world  danger
 due to asterodis "shadows" may offer a dramatic probe
for the international community
to judge the fragility of life. Such dangers call for stronger cooperation
in prospective of a solid peace among people able to preserve life from within
 or outside challenges.

\section*{Acknowledgement}
I wish to thank Dr. R. Conversano for help and discussion.

\end{document}